
\font\llbf=cmbx10 scaled\magstep2
\font\lbf=cmbx10 scaled\magstep1
\def\sq{\hfill \vbox {\hrule width3.2mm \hbox {\vrule height3mm \hskip 2.9mm
                      \vrule height3mm} \hrule width3.2mm }}

\def\ua{\underline a \,}
\def\ub{\underline b \,}
\def\uc{\underline c \,}
\def\ud{\underline d \,}

\def\uA{\underline A \,}
\def\uB{\underline B \,}
\def\uC{\underline C \,}
\def\uD{\underline D \,}

\baselineskip 14pt plus 2pt
\nopagenumbers

\centerline{\llbf Quasi-local energy-momentum and two-surface}
\centerline{\llbf characterization of the}
\centerline{\llbf pp-wave spacetimes}
\bigskip
\centerline{\bf L\'aszl\'o B. Szabados}
\centerline{Research Institute for Particle and Nuclear Physics}
\centerline{H-1525 Budapest 114, P.O.Box 49, Hungary}
\centerline{E-mail: lbszab@rmki.kfki.hu}
\vskip 1.5truecm

\noindent
In the present paper the determination of the {\it pp}-wave metric form
the geometry of certain spacelike two-surfaces is considered. It has been
shown that the vanishing of the Dougan--Mason quasi-local mass $m_{\$}$,
associated with the smooth boundary $\$:=\partial\Sigma\approx S^2$ of
a spacelike hypersurface $\Sigma$, is equivalent to the statement that
the Cauchy development $D(\Sigma)$ is of a {\it pp}-wave type geometry
with pure radiation, provided the ingoing null normals are not diverging
on $\$ $ and the dominant energy condition holds on $D(\Sigma)$. The
metric on $D(\Sigma)$ itself, however, has not been determined.
Here, assuming that the matter is a zero-rest-mass-field, it is shown that
both the matter field and the {\it pp}-wave metric of $D(\Sigma)$ are
completely determined by the value of the zero-rest-mass-field on $\$ $
and the two dimensional Sen--geometry of $\$ $ provided a convexity
condition, slightly stronger than above, holds. Thus the {\it pp}-waves
can be characterized not only by the usual Cauchy data on a {\it three}
dimensional $\Sigma$ but by data on its {\it two} dimensional boundary
$\$ $ too. In addition, it is shown that the Ludvigsen--Vickers quasi-local
angular momentum of axially symmetric {\it pp}-wave geometries has the
familiar properties known for pure (matter) radiation. \par
\bigskip
\bigskip
\vfill \eject

\pageno=1
\footline={\hss\tenrm\folio\hss}

\noindent
{\lbf 1. Introduction}\par
\bigskip
\noindent
The present paper is the third in a series on the 2 dimensional
Sen connection and its applications in general relativity. In the first
[1] a covariant spinor formalism was developed which is the two dimensional
counterpart of the usual (3 dimensional) Sen geometry. In the second paper
[2] this formalism was used to find the `most natural' spinor propagation
law needed in the quasi-local energy-momentum expressions based on the
Nester--Witten form. It turned out that the quasi-local energy-momentum
that the two dimensional Sen operator determines is precisely the
Dougan--Mason energy-momentum [3]. To characterize the zero-mass and zero
energy-momentum spacetimes we proved the following theorem [2,4]:\par

\bigskip
\noindent
{\bf Theorem:} Let $\Sigma$ be a spacelike hypersurface, its boundary,
   $\$:=\partial\Sigma$, be a smooth topological 2-sphere, and suppose
   that the dominant energy condition is satisfied on the Cauchy
   development $D(\Sigma)$ of $\Sigma$. Suppose that the ingoing null
   normals to $\$ $ are not diverging on $\$ $ (or, in other words, the
   GHP spin coeffitient $\rho^\prime$ is nonnegative) and $\$ $ is generic
   (i.e. there exist precisely two linearly independent spinor fields that
   are anti-holomorphic with respect to the 2 dimensional Sen connection
   and span the spin space at each point of $\$ $).
   Then the following pairs of statements are equivalent:\par
\item{1.} the Dougan--Mason quasi-local mass, associated with $\$ $, is
   zero (the quasi-local energy-momentum is zero),
\item{2.} $D(\Sigma)$ is a {\it pp}-wave spacetime with pure radiation
   ($D(\Sigma)$ is flat),
\item{3.} there exists a Sen--constant spinor field (two Sen--constant
   spinor fields) on $\$ $.\par
\bigskip

\noindent In addition to the characterization of the zero-mass and zero
energy-momentum spacetimes the equivalence of 2. and 3. shows that
gravity, together with matter fields satisfying the dominant energy
condition, is so `rigid' a system that the information that $D(\Sigma)$ is
flat/{\it pp}-wave with pure radiation is completely encoded not only into
the Cauchy data on the {\it three} dimensional $\Sigma$, but into the
Sen-geometry of the {\it two} dimensional $\$ $ too. However, while this
theorem tells us in the zero energy-momentum case what the {\it metric} of
$D(\Sigma)$ is, that is flat, in the zero-mass case we know only the {\it
class} of the metric of $D(\Sigma)$: that belongs to the class of the {\it
pp}-wave metrics with pure radiation. One may therefore ask whether all the
information on the metric of a {\it pp}-wave Cauchy development is encoded
into the Sen-geometry of $\$ $. The condition $\rho^\prime\geq 0$ is
usually interpreted as some weak form of the convexity of $\$ $ [3]. However
this condition corresponds to the non-negativity of one of the two {\it
mean} curvatures, while in the theory of surfaces [5,6] the convexity is
defined by the positivity of the {\it Gauss} curvature. Here we show that
the full information on the geometry of a {\it pp}-wave Cauchy development
$D(\Sigma)$ is in fact encoded into the Sen--geometry of $\$ $, provided
certain generalization of the convexity condition of the theory of
surfaces for curved four dimensional embedding geometries holds.\par
        In the first section we review the 2 dimensional Sen--geometry
and define the holomorphic/anti-holom\-orph\-ic spinor fields on $\$ $.
Such a geometry is a quadruple $(\$,\varepsilon_{AB},\gamma_{AB},
\Delta_e)$, where $\varepsilon_{AB}$ is a symplectic and $\gamma_{AB}$ is
a complex metric on the spin spaces and $\Delta_e$ is the covariant
derivation with respect to a Sen--connection. Apart from a common
conformal factor of the two metrics $(\$,\varepsilon
_{AB},\gamma_{AB})$ is analogous to the so-called universal structure of
the geometry of null infinity while $\Delta_e$ represents the so-called
first order structure on $\$ $. Although the 2 dimensional Sen
connections can be introduced as connections on certain SL(2,{\bf
C})-principle bundle (or on an associated complex vector bundle of rank 2)
over a 2 dimensional orientable Riemannian manifold $(\$ ,q_{ab})$ as a
base space [7], for the sake of simplicity we consider the Sen connection
as a structure derived from an imbedding of $\$ $ into a 4 dimensional
Lorentzian spacetime. (The 3 dimensional Sen connection on a 3 dimensional
Riemannian manifold $(\Sigma,h_{ab})$ was introduced in a similar
abstract way without any imbedding only recently [8].) Then in section
3 the Sen--geometry of those 2-surfaces are investigated which are
homeomorphic to $S^2$ and admit a Sen--constant spinor field $\lambda_A$.
We find a function, $\Phi$, describing the deviation of the holomorphic
and anti-holomorphic spin frames from each other, which plays the role
of a potential for the Sen--curvature. In the fourth section we determine
the geometry of the zero-mass Cauchy developments, specified by the
Theorem above, in terms of the 2 dimensional Sen--geometry of $\$ $. A
convexity condition for $\$ $ is found which ensures that both $\Sigma$
and the whole Cauchy development are topologically trivial. This convexity
condition is slightly stronger than that of Dougan and Mason. Then it is
shown that, for any given $(\$,\varepsilon_{AB},\gamma_{AB},\lambda_A)$,
the spacetime curvature and the matter radiation in $D(\Sigma)$ are
completely determined by two complex functions on $\$ $, the potential
$\Phi$ and the value of the matter radiation provided our convexity
condition holds. Then the metric of $D(\Sigma)$ will be determined. If
there is a nonzero constant null vector field in the spacetime then it is
relatively easy to find a local coordinate system $(v,\zeta,\bar\zeta,u)$
(`canonical coordinates') in which the line element contains merely a real
unknown function, $H(\zeta,\bar\zeta,u)$, and Einstein's equations reduce
to a single Poisson equation for $H$ [9]. What we should do here is however
to solve a {\it boundary value problem} for the metric of $D(\Sigma)$ if
the Sen geometry of $\$ $ is fixed. In its spirit our treatment of the
{\it pp}-waves is similar to that of Aichelburg [10]. He also considered
the metric as a solution of a boundary value problem for the Poisson
equation above. However the 2-surface in his treatment was assumed to be
in a hypersurface $v={\rm const}$ of the domain of the canonical coordinate
system. Since a general 2-surface does not lie in such a hypersurface we
should find a weaker coordinate condition. We prove that, for any given
$\$ $, there is a coordinate system (`holomorphic coordinates') whose
domain contains $\$ $ and in these coordinates the line element contains a
real function $H(\zeta,\bar\zeta,u)$ and a complex holomorphic function
$G(\zeta,u)$ and they are determined by the Sen geometry of $\$ $ and the
value of the zero-rest-mass-field on $\$ $. Finally, in section 5, we
discuss the consequences of this result in finding the `correct' quasi-local
angular momentum, the quasi-local radiative modes of gravity and the
possibility of a `quasi-local quantization' of fields and gravity. \par
        Our notations and conventions are the same that used in [1,2]. In
particular, the signature is (+ -- -- --), the curvature and Ricci tensors
are defined by $(\nabla_a\nabla_b-\nabla_b\nabla_a)X^e=:-R^e{}_{fab}X^f$
and $R_{bd}:=R^a{}_{bad}$, respectively, and Einstein's equations then take
the form $G_{ab}=-\kappa T_{ab}$. Throughout this paper the abstract index
formalism [11] will be used, and the concrete indices taking numerical
values will be underlined, e.g. ${\ua}$=0,...,3 and ${\uA}$=0,1.\par
\bigskip
\bigskip

\noindent
{\lbf 2. Review of the 2 dimensional Sen connection}\par
\bigskip

\noindent Let $\$ $ be a smooth connected two dimensional orientable
spacelike submanifold, $t^a$ and $v^a$ future directed unit timelike and
spacelike normals to $\$ $ orthogonal to each other, respectively, and
define $\gamma^A{}_B:=2t^{AR^\prime}v_{BR^\prime}$. (If $\$ $ is the
boundary of some 3-submanifold, $\$=\partial\Sigma$, then $v^a$ will be
assumed to be outward directed from $\Sigma$.) $\gamma^A{}_B$ is
invariant with respect to the conformal rescalings of the spacetime
metric and the `boost gauge transformation' $(t^a,v^a)\mapsto$ $(t^a
\cosh\eta+v^a\sinh\eta,t^a\sinh\eta+v^a\cosh\eta)$. $\gamma^A{}_B$
characterizes the algebraic and conformal properties of $\$ $, since: a.)

$$
\gamma^A{}_A=0,\hskip 20pt \gamma^A{}_B\gamma^B{}_C=\delta^A_C.\eqno(2.1)
$$
\noindent
Thus the eigenvalues of $\gamma^A{}_B$ are $\pm1$ and the corresponding
eigenspinors, defined by $\gamma^A{}_B\iota^B=\iota^A$ and $\gamma^A{}_B
o^B=-o^A$ and normalized by $o_R\iota^R=1$, form a GHP spinor dyad on
$\$ $ [12]. The natural projection $T_pM\rightarrow T_p\$ $, $p\in\$ $,
of the tangent spaces is given by $\Pi^a_b:=\delta^a_b-t^at_b+v^av_b=$
${1\over2}(\delta^A_B\delta^{A^\prime}_{B^\prime}-\gamma^A{}_B\bar
\gamma^{A^\prime}{}_{B^\prime})$. The induced volume form on $\$ $ is
$\varepsilon_{cd}:=t^av^b\varepsilon_{abcd}={{\rm i}\over2}
(\varepsilon_{C^\prime D^\prime}\gamma_{CD}-\varepsilon_{CD}\bar\gamma_{
C^\prime D^\prime})$. $\gamma_{AB}$ can also be considered as a complex
metric on the spin spaces.
b.) $\$ $, together with the induced metric $q_{ab}:=\Pi^e_a\Pi^f_bg_{ef}$,
is a Riemannian manifold whose conformal structure is equivalent to a
complex structure on $\$ $. The projection of the complexified tangent
spaces of $\$ $ to the subspaces of the (1,0) and (0,1) type vectors [7]
are $\pi^{-a}{}_b:=\pi^{-A}{}_B\bar\pi^{+A^\prime}{}_{B^\prime}$ and
$\pi^{+a}{}_b:=\pi^{+A}{}_B\bar\pi^{-A^\prime}{}_{B^\prime}$,
respectively, where $\pi^{\pm A}{}_B:={1\over2}\bigl(\delta^A_B\pm
\gamma^A{}_B\bigr)$ are the projections of the spin space to the subspace
of the $\pm1$ eigenspinors, respectively. $m^a:=o^A\bar\iota^{A^\prime}$
and $\bar m^a:=\bar o^{A^\prime}\iota^A$ are (1,0) and (0,1) type vectors,
respectively.\par
     The two extrinsic curvatures, $\tau_{ab}:=\Pi^e_a\Pi^f_b\nabla_et_f$
and $\nu_{ab}:=\Pi^e_a\Pi^f_b\nabla_ev_f$, can be given in a boost gauge
independent manner by $Q^e{}_{ab}:=-\Pi^e_k\Pi^f_a\nabla_f\Pi^k_b=\tau^e
{}_at_b-\nu^e{}_av_b$. Thus $Q_{eab}=Q_{(ea)b}$ and the expansion tensor
of the out and ingoing null geodesics orthogonal to $\$ $ are $\theta_{ab}
:=\Pi^e_a\Pi^f_b\nabla_el_f=Q_{abk}l^k$ and $\theta^\prime_{ab}:=\Pi^e_a
\Pi^f_b\nabla_en_f=Q_{abk}n^k$, respectively, where $l^a:=o^A\bar
o^{A^\prime}$ and $n^a:=\iota^A\bar\iota^{A^\prime}$. The corresponding
mean curvatures are $q^{ab}\theta_{ab}=-2\rho$ and $q^{ab}\theta^\prime
_{ab}=-2\rho^\prime$; and let us define $k:=\det\Vert\theta^a{}_b\Vert=
{1\over2}(\theta_{ab}\theta_{cd}-\theta_{ac}\theta_{bd})q^{ab}q^{cd}$ and
$k^\prime:=\det\Vert\theta^{\prime a}{}_b\Vert={1\over2}
(\theta^\prime_{ab}\theta^\prime_{cd}-\theta^\prime_{ac}\theta^\prime_{bd})
q^{ab}q^{cd}$. In the theory of surfaces [5,6], when the
imbedding geometry is a flat 3 dimensional Riemannian manifold, we
have only one normal and, because of the Gauss equation, $k$ reduces to
the (real) Gauss curvature $K$ of $\$ $. If however the imbedding geometry
is curved then $k$ and $K$ do not coincide, furthermore in four dimensional
$M$ the relationship between $k$, $k^\prime$ and $K$ is much more
complicated since the Gauss equation gives ${1\over2}R_{abcd}q^{ac}q^{bd}=
K+{1\over2}(\theta_{ac}\theta^\prime_{bd}+\theta^\prime_{ac}\theta_{bd}-
\theta_{ad}\theta^\prime_{bc}-\theta^\prime_{ad}\theta_{bc})q^{ac}q^{bd}$.
In the theory of surfaces the convexity of a 2-surface in a flat 3-space
is defined by the positivity of $K$. By the Gauss equation this definition
is equivalent to the positivity of the principle curvature of the curves in
the 2-surface. In higher dimensional and/or curved imbedding geometries
these definitions are not equivalent. Our convexity condition that will be
used in section 4 is formulated in terms of $k$ and $k^\prime$, or in
other words in terms of the principal curvature of the curves in $\$ $.
\par
       The two dimensional Sen operator is defined by $\Delta_e:=\Pi^f_e
\nabla_f$. The commutator of two Sen operators:

$$\eqalignno{
\bigl(\Delta_c\Delta_d-\Delta_d\Delta_c\bigr)\phi&=-2Q^e{}_{[cd]}\Delta_e
   \phi&(2.2)\cr
\bigl(\Delta_c\Delta_d-\Delta_d\Delta_c\bigr)\xi^A&=-2Q^e{}_{[cd]}\Delta_e
   \xi^A-R^A{}_{Bef}\Pi^e_c\Pi^f_d\xi^B.&(2.3)\cr}
$$

\noindent
The curvature of $\Delta_e$ is therefore the pull back to $\$ $
of the anti-self-dual part of the spacetime curvature: $F^A{}_{Bcd}:=R^A
{}_{Bef}\Pi^e_c\Pi^f_d$; while its `torsion', $T^e{}_{cd}:=2Q^e{}_{[cd]}$,
is determined by the extrinsic curvatures. Expressing $R_{ABcd}$ by the
Weyl and Ricci spinors and the $\Lambda$ scalar $F_{ABcd}$ can be
reexpressed as

$$
F_{ABcd}=-{{\rm i}\over2}\Bigl(\psi_{ABEF}\gamma^{EF}-
   \phi_{ABE^\prime F^\prime}\bar\gamma^{E^\prime F^\prime}+
   2\Lambda\gamma_{AB}\Bigr)\varepsilon_{cd}.\eqno(2.4)
$$

\noindent
The spinor field $\xi^{A...A^\prime...}_{B...B^\prime...}$ will be called
holomorphic/anti-holomorphic if $\pi^{\pm e}{}_f\Delta_e\xi^{...}_{...}=0$.
The quantity $Q^A{}_{eB}:={1\over2}\Delta_e\gamma^A{}_K\gamma^K{}_B$
measures the `non-$\gamma_{AB}$-metricity' of the Sen operator, and the
extrinsic curvature tensor $Q^e{}_{ab}$ can be reexpressed by $Q^A{}_{cB}$
too:

$$
Q^e{}_{ab}={1\over2}\bigl(\delta^{E^\prime}_{B^\prime}Q^E{}_{aB}+
       \delta^E_B\bar Q^{E^\prime}{}_{aB^\prime}+Q^E{}_{aR}\gamma^R{}_B
       \bar\gamma^{E^\prime}{}_{B^\prime}+\bar Q^{E^\prime}{}_{aR^\prime}
	\gamma^E{}_B\bigr).\eqno(2.5)
$$
\noindent
By (2.5) $Q_{AeB}$ is just the anti-self-dual part of the `torsion':
$T_{EE^\prime AA^\prime BB^\prime}=-\bigl(\varepsilon_{A^\prime
B^\prime}Q_{AEE^\prime B}+\varepsilon_{AB}\bar Q_{A^\prime EE^\prime
B^\prime}\bigr)$, furthermore the GHP spin coeffitients $\rho$, $\sigma$
and $\rho^\prime$, $\sigma^\prime$ can also be expressed by $Q_{ARR^\prime
B}$ too. For example $\rho=o^Ao^B\iota^R\bar o^{R^\prime} Q_{ARR^\prime B}$
and $\rho^\prime=-\iota^A\iota^Bo^R\bar\iota^{R^\prime}Q_{ARR^\prime B}$.
\par
      The induced spin connection is defined by

$$
\delta_e\lambda^A:=\Delta_e\lambda^A-Q^A{}_{eB}\lambda^B,\eqno(2.6)
$$

\noindent which for surface tensors is precisely the induced Levi-Civit\`a
covariant differentiation. $\delta_e$ annihilates both $\varepsilon_{AB}$
and $\gamma_{AB}$, and its curvature can be defined by ${}^{\$}R^A{}_{Bcd}
\xi^B:=-(\delta_c\delta_d-\delta_d\delta_c)\xi^A$. Under the boost gauge
transformation the 1-form field $A_e:=\Pi^f_e\nabla_ft_kv^k$ transforms
as an SO(1,1) gauge field: $A_e\mapsto A_e-\Pi^f_e\nabla_f\eta$. By means
of $A_e$ and the curvature scalar ${}^{\$}R$ of the Levi-Civit\`a
connection of $\$ $ ${}^{\$}R_{ABcd}$ can be reexpressed as

$$
{}^{\$}R_{ABcd}=-{1\over2}\gamma_{AB}\Bigl(\bigl(\delta_cA_d-\delta_dA_c
\bigr)-{{}^{\$}R\over4}\bigl(\varepsilon_{C^\prime D^\prime}\gamma_{CD}-
\varepsilon_{CD}\bar\gamma_{C^\prime D^\prime}\bigr)\Bigr).\eqno(2.7)
$$

\noindent
Its imaginary part is the curvature of the usual Levi-Livit\`a (SO(2)--)
connection, while its real part is the curvature of the SO(1,1)--gauge
field $A_e$. With this extension of $\delta_e$ from surface tensors to
spinors we have extended the Levi-Civit\`a covariant differentiation
$\delta_e$ to arbitrary tensors. \par
    To summarize, by a 2 dimensional Sen geometry we mean a quadruple
$(\$,\varepsilon_{AB},\gamma_{AB},\Delta_e)$, where $\varepsilon_{AB}$ is
a symplectic and $\gamma_{AB}$ is a complex metric on a complex vector
bundle ${\bf S}^A(\$)$ of rank 2 over $\$ $ and $\Delta_e$ is a covariant
derivation on ${\bf S}^A(\$)$ such that i. $\gamma_{AB}$ satisfies (2.1);
ii. the complexified tangent bundle $T^{\bf C}\$ $ of $\$ $ is isomorphic
to the Whitney sum of the bundles of the elements $o^A\bar\iota^{A^\prime}$
and $\iota^A\bar o^{A^\prime}$, respectively; iii. $\Delta_e$ annihilates
$\varepsilon_{AB}$; iv. the tensor $Q_{eab}$ defined by the non-$\gamma
$-metricity of $\Delta_e$ according to (2.5) is symmetric in $ea$ and,
finally, v. the derivation $\delta_e$ defined by (2.6) is the (symmetric)
Levi-Civit\`a covariant derivation on $T\$ $ determined by $q_{ab}:={1
\over2}(\varepsilon_{AB}\varepsilon_{A^\prime B^\prime}-\gamma_{AB}\bar
\gamma_{A^\prime B^\prime})$. \par

\bigskip
\bigskip
\noindent
{\lbf 3. Sen--constant spinor fields}\par
\bigskip

\noindent Let $\lambda_R$ be a smooth spinor field on $\$ $ which is
constant with respect to the 2 dimensional Sen connection: $\Delta_e
\lambda_R=0$. First we prove the following lemma, which is the 2
dimensional counterpart of Lemma 2 of [13]:\par
\bigskip
\noindent
{\bf Lemma 3.1} If $\lambda_R$ is Sen--constant on $\$ $ then either it is
identically zero or nowhere zero on $\$ $.\par
\medskip
\noindent {\it Proof:} Let $t^{AA^\prime}$ be any positive definite
hermitian inner product on the spinor spaces, e.g. a future directed
timelike unit normal to $\$ $, and define $h_{ab}:=g_{ab}-t_at_b$.
$h_{ab}$ is a negative definite metric on the 3 dimensional subspaces of
vectors orthogonal to $t^a$. Let $\gamma:[a,b]\rightarrow \$ $ be any
smooth curve and let $X^a$ be its unit tangent. Then since $\lambda_R$ is
Sen--constant we have

$$\eqalign{
\vert X^e\delta_e\bigl(t^{RR^\prime}\lambda_R\bar\lambda_{R^\prime}\bigr)
\vert&=\vert\bigl(X^e\Delta_et^a\bigr)h_{ab}\bigl(h^{bc}\lambda_C\bar
\lambda_{C^\prime}\bigr)\vert\leq\vert h_{ab}\bigl(X^e\Delta_et^a\bigr)
\bigl(X^f\Delta_ft^b\bigr)\vert^{1\over2} \vert h_{ab}\lambda^A
\bar\lambda^{A^\prime}\lambda^B\bar\lambda^{B^\prime}\vert^{1\over2}=\cr
&=\Vert X^e\Delta_et_a\Vert \bigl(t^{RR^\prime}\lambda_R\bar\lambda
_{R^\prime}\bigr);\cr}
$$

\noindent i.e. $\vert {d\over ds}(t^{RR^\prime}\lambda_R\bar\lambda
_{R^\prime})\vert\leq\Vert X^e\Delta_et_a\Vert (t^{RR^\prime}\lambda_R
\bar\lambda_{R^\prime})$, where $s$ is the arch length parameter on
$\gamma$ and $\Vert Y_a\Vert$ is the norm of $Y_e$ in the metric $h_{ab}$.
{}From now on the proof is the same that of Lemma 2 of [13]: If $C\geq0$
such that $C\geq\max\{\Vert X^e\Delta_et_a\Vert(\gamma(s))\vert s\in [a,b]
\}$ then $\vert{d\over ds}(t^{RR^\prime}\lambda_R\bar\lambda_{R^\prime})
\vert\leq C\vert t^{RR^\prime}\lambda_R\bar\lambda_{R^\prime}\vert$, $s\in
[a,b]$. If however $f(s)$ is any $C^1$ function on $[a,b]$ satisfying
$\vert{d\over ds}f(s)\vert\leq C\vert f(s)\vert$ and $f(s_0)>0$ for some
$s_0\in [a,b]$ then $f$ is positive on the whole interval $[a,b]$. Thus if
$\lambda_R$ is zero at a point $p$ of $\$ $ then it must be zero on any
finite piece of any smooth curve through $p$, and therefore on the whole
$\$ $.
\hfill{\sq} \par
\bigskip
\noindent
In [2] we showed that there are at least two linearly independent
holomorphic and two linearly independent anti-holomorphic spinor fields on
$\$ $ provided $\$ $ is homeomorphic to $S^2$. A Sen--constant spinor
field is holomorphic and anti-holomorphic at the same time, and hence one
of the certainly existing two anti-holomorphic and two holomorphic spinor
fields can be chosen to be the constant spinor field $\lambda_R$. Let $\mu
_R$ and $\nu_R$ be the other, nonconstant anti-holomorphic and holomorphic
spinor fields, respectively. The inner products, $\lambda_R\mu^R$ and
$\lambda_R\nu^R$, are always constant on $\$ $, and they are zero if and
only if both $\lambda_R$ and $\mu_R$, and both $\lambda_R$ and $\nu_R$
have a zero [2]. From Lemma 3.1 we can see, however, that $\lambda_R$
cannot have a zero, and hence both $\lambda_R\mu^R$ and $\lambda_R\nu^R$
are nonzero and can be, and in fact will be chosen to be unity. Therefore,
in the terminology of [2], a topological 2-sphere $\$ $ admitting a
constant spinor field is generic; i.e. there are {\it precisely} two
linearly independent anti-holomorphic and two linearly independent
holomorphic spinor fields and both pairs $\lambda_R$, $\mu_R$ and
$\lambda_R$, $\nu_R$ span the spin space at each point of $\$ $. Obviously,
$\mu_R$ and $\nu_R$ are unique up to a complex number times
$\lambda_R$. In the rest of the present paper we assume that $\$ $ is
homeomorphic to $S^2$. \par
      Let us define $\Phi:=\mu_R\nu^R$. Then $\mu_R-\nu_R=\Phi\lambda_R$;
i.e. the difference of the nonconstant anti-holomorphic and holomorphic
spinor fields is proportional to the constant spinor field and the factor
of proportionality is just $\Phi$. There is another interpretation of
$\Phi$: If we consider $(\lambda_R,\mu_R)$ and $(\lambda_R,\nu_R)$ as
two normalized spinor dyads then by

$$
\bigl(\lambda_R,\nu_R\bigr)=\bigl(\lambda_R,\mu_R\bigr)\left(\matrix{
1&-\Phi\cr 0&1\cr}\right) \eqno(3.2)
$$

\noindent
$\Phi$ is the globally well defined $\$ $-dependent parameter of the
SL(2,{\bf C}) spin transformation between the anti-holomorphic and the
holomorphic spin frames. Or, in other words, $\Phi$ measures the deviation
of the holomorphic and anti-holomorphic spin frames from each other.
Obviously $\Phi$ is unique up to an additive constant and at a given
point $\Phi$ can be taken to be zero by a constant
SL(2,{\bf C})-transformation. Furthermore the following statements are
equivalent: 1. $\Phi$ is constant; 2. $\mu_R$ is Sen--constant; 3.
$\nu_R$ is Sen--constant; 4. under the conditions of the Theorem of the
introduction $D(\Sigma)$ is flat. \par
      Let us fix a globally defined smooth spinor field $\omega_R$ such
that $\lambda_R\omega^R=1$. Then since $\lambda_R$ is Sen--constant,
$\Delta_e$ can be specified by the globally defined complex 1-form
$\Gamma_a:=\omega^R\Delta_a\omega_R$ on $\$ $. Since $(\lambda_R,
\omega_R)$ is a spin frame $\mu_R=\omega_R+\alpha\lambda_R$ and $\nu_R=
\omega_R+\beta\lambda_R$ hold for some complex functions $\alpha$ and
$\beta$; and hence $\Phi=\alpha-\beta$. Then since $\lambda_R$ is constant,
$\nu_R$ is holomorphic and $\mu_R$ is anti-holomorphic, the two globally
defined complex valued 1-forms

$$\eqalign{
\Gamma^+_a&:=\pi^{+b}{}_a\bigl(\Delta_b\mu_R\bigr)\mu^R=\Gamma_a+
  \delta_a\alpha=\hskip 7pt \pi^{+b}{}_a\delta_b\Phi\cr
\Gamma^-_a&:=\pi^{-b}{}_a\bigl(\Delta_b\nu_R\bigr)\nu^R=\Gamma_a+
  \delta_a\beta=-\pi^{-b}{}_a\delta_b\Phi\cr}\eqno(3.3)
$$
\noindent
represent the connection coeffitients of the Sen connection in the
anti-holomorphic and holomorphic frames, respectively. They have only
one nonzero component, the other is zero because of the special
holomorphic/\-anti-holomorphic `gauge-choice'. If $\Gamma^+_a=\pi^{+b}{}_a
\delta_b\Phi=\pi^{+b}{}_a\delta_b\Phi^\prime$ then $\pi^{+b}{}_a\delta_b
(\Phi-\Phi^\prime)=0$; i.e. $\Phi-\Phi^\prime$ is holomorphic on $\$ $ and
hence constant. Thus $\Phi$, up to a constant, can be recovered from
either $\Gamma^+_a$ or $\Gamma^-_a$, and hence, apart from constant
SL(2,{\bf C}) transformations, the anti-holomorphic frame and $\Gamma^+_a$
determine the holomorphic frame. Therefore for any given $(\$,\varepsilon
_{AB},\gamma_{AB})$ and $\lambda_R$ there is a 1--1 correspondence between
the functions $\Phi$ modulo constants and the gauge equivalence classes of
the 2 dimensional Sen--connections admitting $\lambda_A$ as a constant
spinor field.\par
     Next calculate the curvature, applying the commutator $\Delta_a
\Delta_b-\Delta_b\Delta_a$ to $\lambda^R$, $\mu^R$ and $\nu^R$,
respectively. We obtain from (2.3), (2.6) and (3.3)

$$
\eqalign{\lambda^SF_{SRab}&=0\cr
\mu^SF_{SRab}&=-\lambda_R\bigl(\delta_a\Gamma^+_b-\delta_b\Gamma^+_a
    \bigr)=-\lambda_R\Bigl(\delta^e_a\pi^{+f}{}_b-
    \delta^e_b\pi^{+f}{}_a\Bigr)\delta_e\delta_f\Phi\cr
\nu^SF_{SRab}&=-\lambda_R\bigl(\delta_a\Gamma^-_b-\delta_b\Gamma^-_b
    \bigr)=\hskip 8pt \lambda_R\Bigl(\delta^e_a\pi^{-f}{}_b-
    \delta^e_b\pi^{-f}{}_a\Bigr)\delta_e\delta_f\Phi.\cr}\eqno(3.4)
$$

\noindent
$\Gamma^\pm_a$ are therefore abelian vector potentials for the curvature
and by $\Gamma^+_a=\pi^{+b}{}_a\delta_b\Phi=(\pi^{+b}{}_a+\pi^{-b}{}_a)
\delta_b\Phi-\pi^{-b}{}_a\delta_b\Phi=\delta_a\Phi+\Gamma^-_a$ they are
gauge-equivalent. The only independent component of the curvature of the
Sen connection is $\mu^R\mu^SF_{RSab}m^a\bar m^b=\nu^R\nu^SF_{RSab}m^a
\bar m^b={1\over2}q^{ab}\delta_a\delta_b\Phi$. Thus $\Phi$ is a potential
for the curvature and by (2.4) we have

$$
q^{ab}\delta_a\delta_b\Phi=\Bigl(\psi_{ABCD}\gamma^{CD}-\phi_{ABC^\prime
D^\prime}\bar\gamma^{C^\prime D^\prime}+2\Lambda\gamma_{AB}\Bigr)\mu^A
\mu^B.\eqno(3.5)
$$

\noindent
Geometrically [7] the constant spinor field $\lambda_R$ defines a
reduction of the pull back to $\$ $ of the SL(2,{\bf C}) spin frame
bundle to a GL(1,{\bf C})$\approx{\bf C}-\{0\}$-principle fibre bundle
over $\$ $ and defines a reduction of the SL(2,{\bf C})-connection to a
GL(1,{\bf C})-connection. \par
      The complex norm $\gamma_{AB}\lambda^A\lambda^B$ of the constant
spinor field can be interpreted tensorially e.g. by the anti-self-dual
simple 2-form $L_{ab}:=\varepsilon_{A^\prime B^\prime}\lambda_A\lambda_B$
since $L^{ab}\varepsilon_{ab}={\rm i}\gamma_{AB}\lambda^A\lambda^B$. Let
us define $z^a:=\Pi^a_b\lambda^B\bar\lambda^{B^\prime}$, whose length is
$\Vert z^e\Vert^2:=-q_{ab}z^az^b={1\over2}\vert\gamma_{AB}\lambda^A
\lambda^B\vert^2$. Thus $\lambda_A$ is null with respect to the complex
metric $\gamma_{AB}$ at $p\in\$ $ iff $z^a(p)=0$. Since $z^a$ is a
continuous vector field on a 2-sphere it must have a zero, and the set
$W:=\{p\in\$\vert z^a(p)\not=0\}$ is obviously open in $\$ $. Since
$\delta_az_b=-Q_{abe}\lambda^E\bar\lambda^{E^\prime}$ and $Q_{abe}=
Q_{(ab)e}$, the 1-form $z_a$ is closed on $\$ $,
and hence by $H^1(S^2)=0$ it is exact: $z_a=\delta_aU$ for some smooth
$U:\$\rightarrow {\bf R}$. For any fixed value $U$ of the function $U$ let
us define $C_U:=\{p\in\$\vert U(p)=U\}$, which are closed sets in $\$ $.
Obviously $z^a=0$ on $\overline{{\rm int}C_U}$, thus $C_U\cap W=(C_U-
{\rm int}C_U)\cap W$. Hence $C_U\cap W$ is a smooth one dimensional
submanifold of $\$ $, i.e. any connected component of $C_U\cap W$ is a
curve $\beta_U(w)$, which is orthogonal to $z^a$. Consequently its tangent
$\dot\beta_U^a$ is proportional to $\varepsilon^{ab}z_b={{\rm i}\over2}
(\delta^{A^\prime}_{B^\prime}\gamma^A{}_B-\delta^A_B\bar\gamma^{A^\prime}
{}_{B^\prime})\lambda^B\bar\lambda^{B^\prime}$: for some real function
$b$ depending on $U$ and the parametrization $w$ $\dot\beta^a_U={1\over
\Vert z^e\Vert}b\varepsilon^{ab}\lambda_B\bar\lambda_{B^\prime}$.
$\beta_U(w)$ is either closed or its endpoints are zeros of $z^a$. \par
\bigskip
\bigskip

\noindent
{\lbf 4. pp-wave Cauchy developments}\par
\bigskip
\noindent
{\bf 4.1 Zero-mass Cauchy developments}\par
\bigskip
\noindent
Let $\lambda_A$ be nonzero constant spinor field on $\$ $. Then by Lemma
3.1 $\$ $ is generic, and hence the Dougan--Mason energy-momentum and mass
are well defined [3,2]: if $\lambda^0_A:=\lambda_A$, $\lambda^1_A:=\mu_A$
then they are $P^{\uA {\uB}^\prime}_{\$}:={{\rm i}\over\kappa}\oint_{\$}
(\overline\lambda^{{\uB}^\prime}_{A^\prime}\Delta_{BB^\prime}\lambda
^{\uA}_A-\overline\lambda^{{\uB}^\prime}_{B^\prime}\Delta_{AA^\prime}
\lambda^{\uA}_B)$ and $m^2_{\$}:=\varepsilon_{{\uA}{\uB}}\varepsilon_{
{\uC}^\prime{\uD}^\prime}P^{{\uA}{\uC}^\prime}_{\$}P^{{\uB}{\uD}^\prime}
_{\$}=2(P^{00^\prime}_{\$}P^{11^\prime}_{\$}-P^{01^\prime}_{\$}P^{1
0^\prime}_{\$})$, respectively. Since $\lambda^0_A$ is constant $P^{0
0^\prime}_{\$}$ is zero. Let $\Sigma$ be a smooth, compact spacelike
hypersurface with the smooth boundary $\$=\partial\Sigma$ and suppose that
$\rho^\prime\geq 0$. Let $t^a$ be the unit future normal to $\Sigma$,
$P^a_b:=\delta^a_b-t^at_b$ the projection to $\Sigma$, $h_{ab}:=P^e_aP^f_b
g_{ef}$ the induced metric, $\varepsilon_{abc}:=t^e\varepsilon_{eabc}$ the
induced volume 3-form, $D_e$ the induced Levi-Civit\`a covariant derivation
and ${\cal D}_e:=P^f_e\nabla_f$ the 3 dimensional Sen operator [13] on
$\Sigma$. Obviously $\Sigma$ fixes a boost gauge on $\$ $ and on the domain
where $\dot\beta^a_U$ is defined the outward unit normal of $\$ $ in
$\Sigma$ can be recovered as $v^a=(\Vert z^e\Vert b)^{-1}\varepsilon^a
{}_{bc}\dot\beta^b_Uz^c$. If $\hat\lambda^{\uA}_R$ is the spinor field on
$\Sigma$ satisfying the Sen--Witten equation ${\cal D}_{R^\prime}{}^R\hat
\lambda^{\uA}_R=0$ with the boundary condition $\pi^{+B}{}_A(\hat\lambda
^{\uA}_B\vert_{\$}-\lambda^{\uA}_B)=0$ then by the Reula--Tod form [14] of
the Sen--identity [13] we obtain for ${\uA}=0$, ${\uA}^\prime=0^\prime$
and for ${\uA}=1$, ${\uA}^\prime=1^\prime$ that

$$
P^{{\uA}{\uA}^\prime}_{\$}={2\over\kappa}\oint_{\$}\rho^\prime\vert\hat
   \lambda^{\uA}_0-\lambda^{\uA}_0\vert^2 d\$+
   {2\over\kappa}\int_\Sigma\Bigl\{-h^{ef}t^{RR^\prime}({\cal D}_e\hat
   \lambda^{\uA}_R)({\cal D}_f\bar{\hat\lambda}{}^{{\uA}^\prime}_{R^\prime})
   -{1\over2}\hat\lambda^{\uA}_A\bar{\hat\lambda}{}^{{\uA}^\prime}
   _{A^\prime}G^{ab}t_b\Bigr\}d\Sigma, \eqno(4.1)
$$

\noindent
where the spinor components are defined by $\lambda^{\uA}_R=:\lambda^{\uA}
_1o_R-\lambda^{\uA}_0\iota_R$. If the dominant energy condition holds on
$\Sigma$ then this implies $P^{11^\prime}_{\$}\geq0$ and $P^{01^\prime}
_{\$}=P^{10^\prime}_{\$}=0$; and, for ${\uA}=0$,
${\uA}^\prime=0^\prime$, that $\hat\lambda_R\mid_{\$}=\lambda_R$ and
${\cal D}_e\hat\lambda_R=0$. By Lemma 2. of [13] $\hat\lambda_R$ is
nowhere zero on $\Sigma$. Foliating $D(\Sigma)$ by a family $\Sigma_t$ of
spacelike Cauchy hypersurfaces (for which $\partial\Sigma_t=\$ $
necessarily holds) and assuming the dominant energy condition to hold on
the whole $D(\Sigma)$, we obtain a smooth nowhere vanishing spinor field
$\hat\lambda_R$ on $D(\Sigma)$. Since $\hat\lambda_A$ is an extension of
$\lambda_A$ form $\$ $ to $D(\Sigma)$, we leave the `hat' and denote this
extension simply by $\lambda_A$ too. In [4] it was shown that $\lambda_R$
is covariantly constant on $D(\Sigma)$, $\nabla_a\lambda_R=0$, and

$$\eqalign{
\psi_{ABCD}&=\psi\lambda_A\lambda_B\lambda_C\lambda_D\cr
\phi_{ABA^\prime B^\prime}&=\phi\lambda_A\lambda_B\overline{\lambda}
  _{A^\prime}\overline{\lambda}_{B^\prime}\cr
\Lambda&=0,\cr}\eqno(4.2)
$$

\noindent
i.e. the Weyl tensor has Petrov N type and the matter is pure radiation
with common wave vector $L^a:=\lambda^A\overline{\lambda}{}^{A^\prime}$.
Here $\psi$ is a complex and $\phi$ is a non-negative real function and
$L^a$ is a covariantly constant nowhere vanishing null vector field on
$D(\Sigma)$. Under the conditions of the Theorem of the introduction a
Sen--constant spinor field $\lambda_R$ on $\$ $ can therefore be extended
into a constant spinor field $\lambda_R$ on $D(\Sigma)$ in a unique way.
As a consequence of (4.2) the Bianchi identities take the following form:

$$
\lambda_A\lambda_B\lambda_C\lambda^D\nabla_{DD^\prime}\psi={1\over3}
\Bigl(\lambda_A\lambda_B\nabla_{CC^\prime}\phi+\lambda_B\lambda_C
\nabla_{AC^\prime}\phi+\lambda_C\lambda_A\nabla_{BC^\prime}\phi\Bigr)
\overline{\lambda}_{D^\prime}\overline{\lambda}{}^{C^\prime}.\eqno(4.3)
$$

\noindent
This implies that $L^e\nabla_e\psi=0=L^e\nabla_e\phi$. By the
definition of $D(\Sigma)$ at each point $p$ of $D(\Sigma)$ the geometrical
properties are uniquely determined by the Cauchy data on $\Sigma$. In
particular, since each $p\in D(\Sigma)$ is on an integral curve of $L^a$
intersecting $\Sigma$ at precisely one point, say $p_0$, the curvature
components $\psi$ and $\phi$ are the same at $p$ and $p_0$. On the other
hand by the normalization $\lambda_R\mu^R=1$ and eq. (4.2) it follows
from (3.5) that on $\$ $

$$
q^{ab}\delta_a\delta_b\Phi=\psi\gamma_{AB}\lambda^A\lambda^B-\phi\bar
\gamma_{A^\prime B^\prime}\bar\lambda^{A^\prime}\bar\lambda^{B^\prime}.
\eqno(4.4)
$$

\noindent
We will show that the curvature on $\Sigma$ is determined, through (4.3)
and (4.4), by $q_{ab}$, $\gamma_{AB}\lambda^A\lambda^B$, $\Phi$ and, if
matter radiation is present, a complex function representing the matter
radiation field on $\$ $. Then we determine the line element of $D(\Sigma)$
in terms of the Sen geometry of $\$ $. First, however, we should clarify
the structure of $\Sigma$.\par
\bigskip

\noindent
{\bf 4.2 The structure of $\Sigma$}\par
\bigskip
\noindent
Since $L_a=\lambda_A\bar\lambda_{A^\prime}$ is constant, it is a gradient:
$L_a=\nabla_a u$ for some smooth $u: {\rm int}D(\Sigma)\rightarrow {\bf
R}$. Furthermore $L^a$ is nowhere vanishing and null on the whole $\Sigma$,
therefore its projection to $\Sigma$, $Z^a:=P^a_bL^b$, is nowhere
vanishing and $z^a=\Pi^a_bZ^b$. Thus $Z^a$ is
orthogonal to $\$ $ at $p\in\$ $ iff $z^a$ has a zero at $p$. $Z_a$ is
also a gradient, $Z_a=D_au$, and hence ${\rm int}\Sigma=\Sigma-\partial
\Sigma$ can be foliated by the 2 dimensional maximal integral submanifolds
$S_u:=\{ q\in{\rm int}\Sigma\mid u(q)=u\}$. Since $\Sigma$ is orientable
the leaves $S_u$ are also orientable, and hence the induced Riemannian
metric defines a complex structure on each $S_u$; i.e. the leaves are
Riemann surfaces. \par
     Let $\{E^a_1,E^a_2,E^a_3\}$ be an orthonormal basis on (an open subset
of) $\Sigma$ such that $E^a_3:={1\over\Vert Z^e\Vert}Z^a$ and $E^a_1$, $E^a
_2$ are also tangent to $\Sigma$, where $\Vert Z^e\Vert^2:=-h_{ef}Z^eZ^f$.
$E^a_3$ is therefore globally well defined but in general $E^a_1$ and
$E^a_2$ are not. The orientation of this basis will be defined by
$\varepsilon_{abc}E^a_1E^b_2E^c_3=1$. Let $N^a$ be the uniquely determined
future null vector field on $\Sigma$ that is orthogonal to $S_u$ and
normalized by $L^aN_a=1$; and let $\sqrt2M^a:=E^a_1-{\rm i}E^a_2$. $M^a$
and $\bar M^a$ are (1,0) and (0,1) type vectors, respectively, in the
complex structure of the Riemann surfaces $S_u$. A local complex
coordinate system $(\zeta,\bar\zeta)$ can always be chosen so that
$M^a=P(\zeta,\bar\zeta,u)({\partial\over\partial\zeta})^a$, where
$P$ is real and positive. $\{ M^a,\bar M^a,E^a_3\}$ is proportional to the
so-called geometric triad of Frauendiener [15]. Let $\chi_{ab}:=P^e_aP^f_b
\nabla_et_f$, the extrinsic curvature of $\Sigma$. Then by taking the
${\cal D}_e$-derivative of $L^a=Z^a+\Vert Z^e\Vert t^a$ we obtain

$$
D_aZ_b=-\Vert Z^e\Vert \chi_{ab}.\eqno(4.5)
$$
\noindent
This implies that the extrinsic curvature $k_{ab}$ of $S_u$ in $\Sigma$ is
just the projection of $\chi_{ab}$ onto $S_u$: $k_{ab}:=(\delta^e_a+E^e_3
E_{3a})(\delta^f_b+E^f_3E_{3b})D_e(-E_{3f})=(\delta^e_a+E^e_3E_{3a})(
\delta^f_b+E^f_3E_{3b})\chi_{ef}$. (It is $-E^a_3$ along that $u$ is {\it
increasing}, thus it is natural to consider $-E^a_3$ as the normal of
$S_u$.) Then by (4.2) and the Gauss equation for the curvature scalar
${}^uR$ of the leaves $S_u$ we have ${}^uR=-2R_{abcd}M^a\bar M^bM^c\bar M^d
=0$; i.e. the Riemann surfaces are locally flat Riemann geometries. The
complex coordinates $(\zeta,\bar\zeta)$ can therefore be chosen so that
$P(\zeta,\bar\zeta,u)=1$ and the remaining allowed transformation of them
is $\zeta=\exp({\rm i}a(u))\zeta^\prime+A(u)$ for real $a(u)$ and complex
$A(u)$.\par
  Next let us clarify the structure of the boundary of the Riemann surfaces
$S_u$. Using the fact that $Z^a$ is a well defined nonzero smooth vector
field on the whole $\Sigma$ one can show that each point $p\in\$ $ belongs
to the closure $\overline{S_u}$ of at most one level surface $S_u$, and
if $p$ does not belong to any $\overline{S_u}$ then $Z^a$ is orthogonal to
$\$ $ at $p$. We can therefore extend $u$ from ${\rm int}\Sigma$ to the
whole $\Sigma$ in the following way: if $p\in\$\cap\overline{S_u}$ for
some $u$ then let $u(p):=u$; and if $p$ does not belong to any $\overline
{S_u}$ then let $\gamma:[0,\varepsilon)\rightarrow\Sigma$ be the integral
curve of $Z^a$ such that $\gamma(0)=p$ and $\dot\gamma^a=\pm Z^a$. If $z$
is the parameter of $\gamma$ then let us define $u(p):=\lim_{z\rightarrow
0}u(\gamma(z))$. This limit always exists since $\pm{{\rm d}u(z)\over{\rm
d}z}=Z^aD_au=h_{ab}Z^aZ^b$, whose right hand side is bounded and $C^\infty$
on $\Sigma$. Next, it is a standard excercise to show that $u$ extended in
this way is a smooth function on the whole $\Sigma$. This result has
important consequences. First, $z_a$ is the gradient of the restriction of
$u$ to $\$ $, i.e. $U=u$ can be chosen. Then since $u:\Sigma\rightarrow
{\bf R}$ is $C^\infty$ and $\Sigma$ is compact $u$ has a minimum $u_-$ and
a maximum $u_+$ somewhere, which must be on $\$ $ because $Z_a$ is nowhere
vanishing. Thus $u\vert_{\$}:\$\rightarrow [u_-,u_+]$ is onto.
Consequently, the topological boundary of any Riemann surface $S_u$, $B_u:=
\$\cap\overline{S_u}$, is nonempty and $B_u=C_u-{\rm int}C_u$. Thus, as we
saw at the end of the previous section, the piece of the topological
boundary $B_u$ lying in the open domain where $z^a\not=0$, i.e. $B_u\cap
W$, consists of smooth curves $\beta_u(w)$ orthogonal to $z^a$. Without
further conditions on the geometry of $\$ $ one cannot say anything about
the structure of $B_u$ at the zeros of $z^a$. The next proposition shows
that certain generalization of the convexity conditions of differential
geometry, however, excludes the strange structures for $B_u$.\par
\bigskip
\noindent{\bf Proposition 4.6} Let $k>0$ and $k^\prime>0$, and the
  ingoing null normals be converging and the outgoing null normals be
  diverging somewhere on $\$ $. Then
  \item{1.} $z^a$ has two isolated zeros, $p_\pm$, for which $u(p_\pm)
            =u_\pm$, $p_\pm$ do not belong to the closure of any $S_u$,
            and the topological boundary $B_u$ of any Riemann surface
             $S_u$, $u\in(u_-,u_+)$, consists of a single smooth closed
             curve $\beta_u(w)$;
  \item{2.} each Riemann surface $S_u$ is homeomorphic to ${\bf R}^2$;
  \item{3.} $\Sigma$ is homeomorphic to the closed three--ball $B^3$. \par

\bigskip
\noindent
{\it Proof:} (1) First, on the contrary, suppose that for some $S_u$ there
is a point $p\in\$\cap\overline{S_u}$ which is a zero of $z^a$. The
normal $-E^a_3$ of $S_u$ can obviously be extended to $\overline{S_u}$,
and $\$ $ and $\overline{S_u}$ are tangent at $p$ to each other: $E^a_3=
\pm v^a$. (Here $v^a$ is the outward directed unit normal to $\$ $ in
$\Sigma$.) Then the position of $\$ $ relative to $\overline{S_u}$ can be
determined by comparing the principle curvature of the curves through $p$
lying in $\$ $ and in $\overline{S_u}$ (see e.g. [5,6]). Let therefore
$X^a\in T_p\$=T_p\overline{S_u}$ be a unit vector and $\gamma$ and
$\gamma_u$ be smooth curves in $\$ $ and $\overline{S_u}$, respectively,
whose tangent at $p$ is $X^a$ and whose principle normal at $p$ is
proportional to $v^a$. Let $\kappa^X$ and $\kappa^X_u$ be the
corresponding principle curvatures at $p$. Let the NP null tetrad
$\{l^a,m^a,\bar m^a,n^a\}$, adapted to $\$ $, be normalized by
$\sqrt2 l^a=t^a+v^a$ and $\sqrt2 n^a=t^a-v^a$. Then a short calculation
shows that the difference of the principle curvatures are determined by
the quadratic form defined by the expansion tensors of $\$ $:

$$\eqalignno{
\kappa^X-\kappa^X_u&=-\Bigl(\Pi^e_a\Pi^f_bD_ev_f-\Pi^e_a\Pi^f_bD_eE_{3f}
\Bigr)X^aX^b=-\sqrt2\theta_{ab}X^aX^b \hskip 10pt {\rm if}\hskip 10pt
E^a_3=v^a&(4.7a)\cr
\kappa^X-\kappa^X_u&=-\Bigl(\Pi^e_a\Pi^f_bD_ev_f+\Pi^e_a\Pi^f_bD_eE_{3f}
\Bigr)X^aX^b=\hskip 7pt \sqrt2\theta^\prime_{ab}X^aX^b \hskip 10pt
{\rm if}\hskip 10pt E^a_3=-v^a.&(4.7b)\cr}
$$

\noindent
By the Cayley--Hamilton equation we have $\theta_{ab}\theta_{ef}q^{ef}=
q_{ab}k+q^{ef}\theta_{ea}\theta_{fb}$, whose right hand side is negative
definite for everywhere positive $k$. Hence for somewhere positive
$\theta_{ef}q^{ef}$ this implies that $\theta_{ab}$ is negative definite;
i.e. $\kappa^X>\kappa^X_u$. However $\kappa^X\leq\kappa^X_u$ must hold, as
otherwise $\overline{S_u}$ would be {\it outwardly} tangent at $p$ to
$\$ $. Similarly the Cayley--Hamilton equation, $k^\prime>0$ on $\$ $ and
$\theta^\prime_{ab}q^{ab}<0$ somewhere on $\$ $ imply the positive
definiteness of $\theta^\prime_{ab}$, which would contradict $\kappa^X\leq
\kappa^X_u$. Thus no point of $\$\cap\overline{S_u}$ can be a zero of
$z^a$ and hence by the argumentation just before the present proposition
each connected component of $B_u$ is a single closed smooth curve. Thus
$z^a$ may have zeros only at the points $a\in\$ $ where $u(a)=u_\pm$.\par
        Let $A$ be the preimage of $u_-$ by $u$; i.e. $A:=\{a\in\$\vert
u(a)=u_-\}$. Obviously $A$ is closed. If $A$ were not connected, say
$A=A^\prime\cup A^{\prime\prime}$ for disjoint nonempty closed sets
$A^\prime$ and $A^{\prime\prime}$, then since $\$ $ is connected,
$u\vert_{\$}:\$\rightarrow{\bf R}$ is smooth and $u_-$ is the minimum
value of $u$, there would be a point $p\in\$-A$ where $u(p)\in(u_-,u_+)$
and $z_a(p)=\delta_a u(p)=0$. This however would contradict the first part
of the present proof, thus $A$ must be connected and, since $\$ $ is a
manifold, path connected. Let $a\in A$ and suppose that $A-\{a\}$ is not
empty. Then there is a series $\{a_n\}$ of points of $A-\{a\}$ converging
to $a$. Let $\gamma_n$ be the geodesic in $\$ $ through $a$ and $a_n$; and
let $X^a_n$ be its unit tangent at $a$. Then there is a unit vector $X^a$
at $a$ such that $\{X^a_n\}$ (or at least a sub-series of it) converges
to $X^a$. Then it is easy to prove that $X^a(\delta_a\delta_bu)=0$ at
$a\in\$ $. Let $\gamma$ be the geodesic in $\$ $ through $a$ with tangent
$X^a$. At the points of $A$ $E^a_3=v^a$, thus if $F_u$ is the 1-parameter
family of local diffeomorphisms generated by $-{1\over\Vert Z^e\Vert^2}Z^a$
then for sufficiently small $\epsilon>0$ $F_\epsilon(A)\subset S_{u_-+
\epsilon}$. Let $\gamma^\epsilon:=F_\epsilon\circ\gamma$, whose tangent
$F_{\epsilon *}(X^a)$ is easily seen to lie in $S_{u_-+\epsilon}$. Finally
let $\bar\gamma^\epsilon$ be the smooth curve in $S_{u_-+\epsilon}$ through
$F_\epsilon(s)$ whose projection along the orbits of $F_u$ to $\$ $ is
$\gamma$. Since in general $\gamma$ does not lie in $A$ $\gamma^\epsilon$
does not coincide with $\bar\gamma^\epsilon$. However the tangent of $\bar
\gamma^\epsilon$ at $F_\epsilon(a)$ is just $F_{\epsilon *}(X^a)$;
furthermore by $X^a(\delta_a\delta_bu)(a)=0$ the principle curvature of
$\gamma^\epsilon$ and $\bar\gamma^\epsilon$ are equal at $a$. But the
$\epsilon\rightarrow0$ limit of the difference of these principle
curvatures is $-\sqrt2 \theta_{ab}X^aX^b>0$, which is a contradiction.
Thus $A$ consists of a single point, or, in other words, $u\vert_{\$}:\$
\rightarrow{\bf R}$ takes its minimum value at a single point $p_-$. By a
similar argumentation, it takes its maximal value at another single point
$p_+$. Since however $z^a$ may have zeros only at the points where $u$
takes its minimal or maximal value, the two zeros of $z^a$ must be $p_\pm$,
which are isolated. What remained to show is that $B_u$ is connected for
any $u\in(u_-,u_+)$.\par
      Let $U$ be any open neighbourhood of $p_-$ in $\Sigma$. Then for
sufficiently small $\epsilon>0$ $\$_{u_-+\epsilon}:=\{p\in\$\vert u(p)<u_-+
\epsilon\}\subset U\cap\$ $ and $\partial\$_{u_-+\epsilon}=B_{u_-+
\epsilon}$, which is connected. Since the only zeros of $z^a$ are $p_{\pm}$,
$-{1\over\Vert z^e\Vert^2}z^a$ defines a 1-parameter family $f_u$ of local
diffeomorphisms of $\$-\{p_-,p_+\}$ onto itself. Since the action of these
diffeomorphisms is transitive and $f_{u-u^\prime}B_{u^\prime}=B_u$, any
$B_u$ can be mapped into $B_{u_-+\epsilon}$. Thus $B_u$ is connected. (See
also [16].)\par
(2) Let $Y^a$ be any smooth vector field on $\Sigma$ which is tangent to
the leaves $S_u$, orthogonal to $B_u$ and nonzero on $\$ $. Since $v^a$
is the unit normal to $\$ $ it is a linear combination of $Y^a$ and $Z^a$
on $\$ $. Then for any pair of smooth functions $\alpha,\beta:\Sigma-
\{p_-,p_+\}\rightarrow {\bf R}$ let us define $\tilde z^a:=\alpha Y^a+
\beta Z^a$. For everywhere nonzero $\beta$ $\tilde z^a$ is nowhere zero
on $\Sigma-\{p_-,p_+\}$, and $\tilde z^aD_au=\beta h_{ab}Z^aZ^b$.
Furthermore if the restriction of $\alpha$ and $\beta$ to $\$ $ are chosen
to satisfy $\Vert Y^e\Vert^2\alpha\mid_{\$}=-(v^eY_e)(v^fZ_f)$ and $\Vert
Z^e\Vert^2\beta\mid_{\$}=\Vert Z^e\Vert^2-(v^eZ_e)^2$, respectively, then
$\tilde z^a\mid_{\$}=z^a$. With this choice $\tilde z^a$ is a nowhere
vanishing extension of $z^a$ to $\Sigma-\{p_-,p_+\}$. Thus if $\tilde F_u$
is the 1 parameter family of local diffeomorphisms of $\Sigma-\{p_-,p_+\}$
generated by $-{1\over\beta\Vert Z^e\Vert^2}\tilde z^a$ then it is
transitive on $\Sigma-\{p_-,p_+\}$ and $\tilde F_{u-u^\prime}S_{u^\prime}=
S_u$. Since for sufficiently small $\epsilon>0$ $S_{u_-+\epsilon}$ is
homeomorphic to ${\bf R}^2$ this implies that $S_u$ is homeomorphic to
${\bf R}^2$ for any $u\in(u_-,u_+)$.\par
   (3) is now a simple consequence of the result that $\Sigma-\{p_-,p_+\}$
is homeomorphic to $B^2\times(u_-,u_+)$ where $B^2$ is the closed 2-ball.
\hfill{\sq}
\par

\bigskip
\noindent
An immediate consequence of this proposition is that the Riemann surfaces
$S_u$ are connected, simply connected subsets of a flat plane and they form
a global foliation of $\Sigma$. (Its lapse $n$ is given by $n^{-1}:=
(-E_3^a)D_au=\Vert Z^e\Vert$.) Consequently the complex null coordinate
system $(\zeta,\bar\zeta)$ is globally defined on the Riemann surfaces
$S_u$ and hence on the whole $\Sigma$. In this coordinate system the
closed curves $\beta_u(w)$ can be given as $(\zeta(w,u),\bar\zeta(w,u),u)$.
If the parameter $w$ is chosen such that $b=b(u)$ is ${1\over2\pi}$ times
the arch length of $\beta_u$ then $(w,u)$, $w\in[0,2\pi)$ $u\in(u_-,u_+)$,
is a coordinate system on $\$-\{p_{\pm}\}\approx S^1\times (u_-,u_+)$. Note
that this coordinate system is determined by the geometry of $\$ $ alone.\par
       Let $I_A$ be the smooth spinor field on $\Sigma$ for which
$\lambda_AI^A=1$ and the complex null vectors $\lambda^A\bar I^{A^\prime}$,
$I^A\bar\lambda^{A^\prime}$ are tangent to the Riemann surfaces $S_u$.
These conditions uniquely determine $I^A$, and $I^A\bar I^{A^\prime}=N^a$,
$\lambda^A\bar I^{A^\prime}=\exp({\rm i}\omega)M^a$ hold for some smooth
function $\omega:\Sigma\rightarrow{\bf R}$. Let $v$ be the affine parameter
along the (maximally extended null geodesic) integral curves of $L^a$
measured from $\Sigma$ and Lie propagate $(\zeta,\bar\zeta)$ along $L^a$.
($u$ is automatically Lie propagated along $L^a$.) Then $D(\Sigma)$ is
covered by the domain of the coordinate system $(v,\zeta,\bar\zeta,u)$.
In this coordinate system the spacetime metric takes the
form $ds^2=2dvdu-2d\zeta d\bar\zeta+ 2(Gd\zeta+\bar Gd\bar\zeta)du +
2Hdu^2$, where $H$ is a real and $G$ is a complex function of $\zeta$,
$\bar\zeta$ and $u$. Since ${\cal D}_a\lambda_R=0$ and $M^a=\exp(-{\rm i}
\omega)\lambda^A\bar I^{A^\prime}$, we have $M^e{\cal D}_a\bar M_e=-{\rm
i}D_a\omega$. Its left hand side can be reexpressed by the Christoffel
symbols of the spacetime metric and we obtain $-{\rm i}D_a\omega=M^e{\cal
D}_a\bar M_e={1\over2}\bigl({\partial G\over\partial\bar\zeta}-{\partial
\bar G\over\partial\zeta}\bigr)D_au.$ This implies that $\omega$ depends
only on $u$, and hence by the allowed transformation of the coordinates
$(\zeta,\bar\zeta)$ by $a(u)=-\omega(u)$ it can be taken zero. Thus in
this coordinate system $M^a=\lambda^A\bar I^{A^\prime}$, $\bar M^a=I^A
\bar\lambda^{A^\prime}$ and

$$
{\partial G\over\partial\bar\zeta}-{\partial\bar G\over\partial\zeta}=0.
\eqno(4.8)
$$
\noindent
In the rest of this paper the complex coordinates will be chosen to
satisfy these properties. Obviously all the spinor and vector fields on
$\Sigma$ can be extended onto the whole domain of the coordinate system
by Lie propagation along $L^a$. \par
\bigskip

\noindent
{\bf 4.3 The curvature and the metric of $D(\Sigma)$}\par
\bigskip
\noindent
Returning to the spinor Bianchi identity let us contract (4.3) with
$I^AI^BI^C\bar I^{D^\prime}$. We obtain

$$
M^a\nabla_a\psi=\bar M^a\nabla_a\phi.\eqno(4.9)
$$

\noindent
If $\phi$ were zero then by (4.9) $\psi$ would be anti-holomorphic on
each Riemann surface $S_u$, and, since topologically $S_u$ is trivial and
its boundary $B_u$ is a single closed curve, $\psi$ on $S_u$ would
completely be determined by its value on $B_u$, and hence by $q_{ab}$,
$\gamma_{AB}\lambda^A\lambda^B$ and $\Phi$ (cf. eq.(4.4)). Similarly,
$\psi$ could be determined from its boundary value for any {\it given}
$\phi$ using e.g. the Green function method of Aichelburg [10]. Since
however (4.9) is only one equation for $\psi$ and $\phi$, it does not
determine them on $\Sigma$ from {\it their boundary value} on $B_u$ unless
the field equation for the matter fields are specified. \par
      For the massless complex scalar field $\varphi$ the energy-momentum
tensor takes the form $T_{ab}=\nabla_{(a}\varphi\nabla_{b)}\bar\varphi
-{1\over2}g_{ab}g^{ef}\nabla_e\varphi\nabla_f\bar\varphi$, thus the
condition $T_{ab}L^b=0$ implies that $\nabla_a\varphi=fL_a$ for some
complex function $f$. Therefore $M^a\nabla_a\varphi=0$ and $\bar M^a
\nabla_a\varphi=0$; i.e. $\varphi$ is constant on each Riemann surface
$S_u$ and hence $\varphi$ on $\Sigma$ is completely determined by its
value on $\$ $. \par
        For the electromagnetic field the field strength is described
by a symmetric spinor $\varphi_{AB}$ defined by $F_{ab}=\varepsilon_{
A^\prime B^\prime}\varphi_{AB}+\varepsilon_{AB}\bar\varphi_{A^\prime
B^\prime}$, Maxwell's source free equations take the form $\nabla
_{A^\prime}{}^A\varphi_{AB}=0$ and the energy-momentum tensor is $T_{ab}=
2\varphi_{AB}\bar\varphi_{A^\prime B^\prime}$. If algebraically $T_{ab}$
is pure radiation with wave vector $L^a$ then $\varphi_{AB}=\varphi
\lambda_A\lambda_B$ for some complex function $\varphi$, the only nonzero
component of the Ricci spinor is $\phi=\kappa\varphi\bar\varphi$ and
Maxwell's equations reduce to $\lambda^A\nabla_{AA^\prime}\varphi=0$.
Contracting it with $\bar I^{A^\prime}$ we obtain $M^a\nabla_a\varphi=0$.
Thus $\varphi$ is anti-holomorphic on each Riemann surface $S_u$ and
therefore both $\varphi$ and $\phi$ are completely determined on $\Sigma$
by the value of $\varphi$ on $\$ $. Physically, $\varphi$ is the
complex combination of the electric and magnetic field strengths defined
by the normal $t^a$ of $\Sigma$: $E_a+{\rm i}B_a=-\varphi\Vert Z^e\Vert
M_a$. This implies that $E^aE_a=B^aB_a$, $E^aB_a=0$ and $Z^a(E_a+{\rm i}
B_a)=0$, which are the characteristic properies of plane electromagnetic
fields with spatial wave vector proportional to $Z^a$. \par
       One can consider general $2s$th order symmetric spinor fields
$\varphi_{AB...D}$ satisfying the zero-rest-mass-field equation
$\nabla_{A^\prime}{}^A\varphi_{AB...D}=0$ with {\it integer} $s$. (For
half-integer $s$ the energy-momentum tensor is not expected to satisfy
the dominant energy condition. In fact, for the Weyl neutrino field
$\varphi_A$ the energy-momentum tensor is $T_{ab}={\rm i}(\bar\varphi_{
A^\prime}\nabla_{BB^\prime}\varphi_A-\varphi_A\nabla_{BB^\prime}\bar
\varphi_{A^\prime}+\bar\varphi_{B^\prime}\nabla_{AA^\prime}\varphi_B-
\varphi_B\nabla_{AA^\prime}\bar\varphi_{B^\prime})$, which does not satisfy
even the weak energy condition.) $\varphi_{AB...D}$ is said to describe a
pure radiation with wave vector $\lambda^A\overline{\lambda}{}^{A^\prime}$
if $\varphi_{AB...D}=\varphi\lambda_A...\lambda_D$; and the energy-momentum
tensor of this radiation field can be {\it defined} by $T_{ab}:=2\varphi
\bar\varphi\lambda_A\lambda_B\overline{\lambda}_{A^\prime}\overline{\lambda
}_{B^\prime}$. Then

$$
M^a\nabla_a\varphi=0;\eqno(4.10)
$$
\noindent
i.e. $\varphi$ is anti-holomorphic on the Riemann surfaces $S_u$ and hence
$\phi$ on $\Sigma$ is determined by the value of $\varphi$ on $\$ $. To
summarize, if we assume that the matter field is a pure radiative massless
complex scalar field or zero-rest-mass-field with integer helicity, then
the field equations imply (4.10) and, for any solution $\varphi$ of (4.10)
on $\Sigma$, the solution of (4.9) for $\psi$ can be specified by the
value of the solution $\psi$ on $\$ $. Therefore, for given $q_{ab}$ and
$\gamma_{AB}\lambda^A\lambda^B$, the curvature on $\Sigma$ (and
consequently along all the integral curves of $L^a$ crossing $\Sigma$) is
determined by the complex functions $\Phi$ and $\varphi\vert_{\$}$. \par
     Finally determine the metric of $D(\Sigma)$ from the data on $\$ $.
The only nonzero components of the curvature are

$$\eqalign{
\phi&=I^AI^BR_{ABcd}N^cM^d={\partial\over\partial\zeta}\bigl(
      {\partial H\over\partial\bar\zeta}-{\partial\bar G\over\partial u}
      \bigr)\cr
\psi&=I^AI^BR_{ABcd}N^c\bar M^d={\partial\over\partial\bar\zeta}\bigl(
      {\partial H\over\partial\bar\zeta}-{\partial\bar G\over\partial u}
      \bigr).\cr}
\eqno(4.11)
$$
\noindent
The integrability condition of (4.11) is just (4.9). To show that the
Sen geometry of $\$ $ determines the metric of $D(\Sigma)$ completely
we should fix a gauge.
For any smooth real valued function $V$ of $\zeta$, $\bar\zeta$ and $u$
the mapping ${\tt V}:(v,\zeta,\bar\zeta,u)\mapsto(v+V(\zeta,\bar\zeta,u),
\zeta,\bar\zeta,u)$ is a smooth diffeomorphism of the domain of the
coordinate system onto itself. Geometrically ${\tt V}$ shifts the zero of
the affine parameter $v$ of $L^a$ by $V$; and let $\Sigma^\prime:={\tt V}
(\Sigma)$. Then the action of ${\tt V}$ on the coordinate vectors is
${\tt V}_*(L^a)=L^a$, $M^{\prime a}:={\tt V}_*(M^a)=M^a+{\partial V\over
\partial\zeta}L^a$, $\bar M^{\prime a}:={\tt V}_*(\bar M^a)=\bar M^a+
{\partial V\over\partial\bar\zeta}L^a$ and $({\partial\over\partial
{u^\prime}})^a:={\tt V}_*(({\partial\over\partial u})^a)=({\partial\over
\partial u})^a+{\partial V\over\partial u}L^a$. The pull back $g^\prime
_{ab}:={\tt V}^*(g_{ab})$ of the spacetime metric is $ds^{\prime 2}=2dudv
-2d\zeta d\bar\zeta+2((G+{\partial V\over\partial\zeta})d\zeta+(\bar G+
{\partial V\over\partial\bar\zeta})d\bar\zeta)du+2(H+{\partial V\over
\partial u})du^2$. Thus in the coordinates $(v,\zeta,\bar\zeta,u)$ the
metrics $g_{ab}$ and $g^\prime_{ab}$ have the same form if $G^\prime:=G+
{\partial V\over\partial\zeta}$ and $H^\prime:=H+{\partial V\over
\partial u}$. Let $N^{\prime a}$ be the future directed null vector field
orthogonal to $M^{\prime a}$ and $\bar M^{\prime a}$, and normalized by
$N^\prime_aL^a=1$. In the new basis $N^{\prime a}$ is given by $N^{\prime
a}=-(H^\prime+\bar G^\prime G^\prime)L^a+({\partial\over\partial{u^\prime}}
)^a+\bar G^\prime M^{\prime a}+G^\prime \bar M^{\prime a}$. Since in
general $\Sigma^\prime$ is {\it not} a Cauchy surface for $D(\Sigma)$ (it
might become timelike or null somewhere and some portions of $\Sigma
^\prime$ may even be outside $D(\Sigma)$), $\Sigma$ is not necessarily
spacelike with respect to $g^\prime_{ab}$. Its normal is proportional to
$N^{\prime a}-(H^\prime+G^\prime\bar G^\prime)L^a$. Thus $\Sigma$ is
spacelike/null/timelike at a point $p\in\Sigma$ if $H^\prime+G^\prime\bar
G^\prime$ is negative/zero/positive there; and define $\Sigma_+$, $\Sigma
_0$ and $\Sigma_-$ the spacelike, null and timelike pieces of $\Sigma$,
respectively. On $\Sigma_\pm$ the length of the normal can be chosen to be
unity: $t^{\prime a}:=\pm{1\over\sqrt{2\vert H^\prime+G^\prime\bar
G^\prime\vert}}(N^{\prime a}-(H^\prime+G^\prime\bar G^\prime)L^a)$, and in
the null case $t^{\prime a}$ will be chosen to be $N^{\prime a}$.
On $\Sigma_\pm$ $P^{\prime a}_b:=\delta^a_b
\mp t^{\prime a}t^\prime_b$ is the projection, and let $Z^{\prime a}:=
P^{\prime a}_bL^b={1\over2}(L^a+{1\over H^\prime+G^\prime\bar G^\prime}
N^{\prime a})$. Its norm is $\Vert Z^{\prime e}\Vert^2:=\mp g^\prime_{ab}
Z^{\prime a}Z^{\prime b}=\mp{1\over2(H^\prime+G^\prime\bar G^\prime)}$.
Let $I^\prime_A$ be the (uniquely determined) spinor field for which
$\lambda_AI^{\prime A}=1$ and $N^\prime_a=I^\prime_A\bar I^\prime_{
A^\prime}$ hold and $I^{\prime A}\bar\lambda^{A^\prime}$, $\lambda
^A\bar I^{\prime A^\prime}$ are tangent to the (flat) 2-surfaces $u={\rm
const}$, $v+V(\zeta,\bar\zeta,u)={\rm const}$. Then it is easy to show
that $I^\prime_A=I_A+({\partial V\over\partial\bar\zeta})\lambda_A$; which
implies that $M^{\prime a}=\lambda^A\bar I^{\prime A^\prime}$, $\bar M^{
\prime a}=I^{\prime A}\bar\lambda^{A^\prime}$ and that $I^\prime _A$ is
Lie propagated along $L^a$. The only nonzero components of the curvature
$R^\prime _{ABcd}$ of $g^\prime_{ab}$ in the basis $(\lambda_A,I_A)$ are
$\phi^\prime:=I^AI^BR^\prime_{ABcd}N^cM^d=I^{\prime A}I^{\prime B}R_{ABcd}
N^{\prime c}M^{\prime d}=I^AI^BR_{ABcd}N^cM^d=\phi$ and $\psi^\prime:=
I^AI^BR^\prime_{ABcd}N^c\bar M^d=\psi$; i.e. (4.11) remains valied for the
primed $H^\prime$ and $G^\prime$ on the right too. By (4.8) there exists a
$V$ for which $G^\prime=0$ (`canonical gauge' [9,10]). In general however
this $V$ is not zero on $\$ $; i.e. such a transformation would {\it
deform the boundary $\$ $} too. Since the boundary is fixed in our problem
and the boundary values for the metric of $D(\Sigma)$ are given on $\$ $,
we should find a weaker gauge condition for $G^\prime$ such that the
corresponding diffeomorphism ${\tt V}$ leaves $\$ $ fixed. \par
    A weaker gauge condition might be the requirement of the holomorphity
of $G^\prime$. The condition of the existence of a transformation of the
form above yielding holomorphic $G^\prime$ is ${\partial^2V\over\partial
\zeta\partial\bar\zeta}=-{\partial G\over\partial\bar\zeta}$. By (4.8)
this is a Poisson equation for $V$ with real source, which always has a
unique solution with the boundary condition $V\vert_{\$}=0$. Thus there
exists a diffeomorphism ${\tt V}$ such that $G^\prime$ in $g^\prime_{ab}
={\tt V}^*(g_{ab})$ is holomorphic (`holomorphic gauge'). In this gauge
(4.11) reduces to $\phi={\partial^2 H^\prime\over\partial\zeta\partial
\bar\zeta}$ and $\psi+{\partial^2\bar G^\prime\over\partial u\partial\bar
\zeta}={\partial^2 H^\prime\over\partial\bar\zeta^2}$. Since however
$G^\prime$ is holomorphic, it is determined by its value on $\$ $; and
hence $\phi$, $\psi$ and the value of $H^\prime$ and $G^\prime$ on $\$ $
determine $H^\prime$ and $G^\prime$ on the whole coordinate domain. What
remained to show is that $H^\prime\vert_{\$}$ and $G^\prime\vert_{\$}$
are determined by the Sen geometry of $\$ $ and the boost gauge on $\$ $
defined by $\Sigma^\prime$. \par
     Since $V\vert_{\$}=0$ the 2-surface coordinate vectors can be
expressed in terms of $M^{\prime a}$, $\bar M^{\prime a}$ and $({\partial
\over\partial{u^\prime}})^a$ too: $({\partial\over\partial u})^a_{\$}=
{\partial\zeta\over\partial u}M^{\prime a}+{\partial\bar\zeta\over\partial
u}\bar M^{\prime a}+({\partial\over\partial {u^\prime}})^a$ and $({\partial
\over\partial w})^a_{\$}={\partial\zeta\over\partial w}M^{\prime a}+
{\partial\bar\zeta\over\partial w}\bar M^{\prime a}$. The first implies
that $G^\prime=\lambda_A\bar I^\prime_{A^\prime}({\partial\over\partial
u})^a_{\$}+{\partial\bar\zeta\over\partial u}$ and $H^\prime=I^\prime_A
\bar I^\prime_{A^\prime}({\partial\over\partial u})^a_{\$}-G^\prime\bar
G^\prime$. Thus we should show that $I^{\prime A}$ is determined by the
Sen geometry and the boost gauge. Since $\gamma_{AB}\lambda^A\lambda^B$
is nonzero on $\$-\{p_\pm\}$ $\lambda_A$ and $\gamma_{AB}\lambda^B$ span
the spin space at each point of $\$-\{p_\pm\}$. Thus by the continuity of
$I^{\prime A}$ and by $\lambda_AI^{\prime A}=1$ the spinor $I^{\prime A}$
on $\$ $ is determined by $\gamma_{AB}\lambda^AI^{\prime B}$. We will show
that $\gamma_{AB}\lambda^AI^{\prime B}$ is determined by the boost gauge
on $\$ $. Since $({\partial\over\partial w})^a_{\$}=\dot\beta_u={{\rm i}
\over2\Vert z^e\Vert}b(\bar\lambda^{A^\prime}\gamma^A{}_B\lambda^B-
\lambda^A\bar\gamma^{A^\prime}{}_{B^\prime}\bar\lambda^{B^\prime})$ we
obtain ${\partial\bar\zeta\over\partial w}=-\dot\beta^a_uM^\prime_a=-{{\rm
i}\over2\Vert z^e\Vert}b\gamma_{AB}\lambda^A\lambda^B$; and by $0=\dot
\beta^a_uN^\prime_a={{\rm i}\over2\Vert z^e\Vert}b(\gamma_{AB}\lambda^A
I^{\prime B}-\bar\gamma_{A^\prime B^\prime}\bar\lambda^{A^\prime}\bar
I^{\prime B^\prime})$ the complex scalar product $\gamma_{AB}\lambda^A
I^{\prime B}$ is {\it real}. Since $z^a={1\over2}(\lambda^A\bar\lambda^{
A^\prime}-\gamma^A{}_B\lambda^B\bar\gamma^{A^\prime}{}_{B^\prime}\bar
\lambda^{B^\prime})$ is $\Vert z^e\Vert$ times the unit vector orthogonal
to $\dot\beta^a_u$ it has the form $z^a=-\Vert z^e\Vert^2(N^{\prime a}+
(H^\prime+G^\prime\bar G^\prime)L^a)+{1\over4}(\bar\gamma_{A^\prime
B^\prime}\bar\lambda^{A^\prime}\bar\lambda^{B^\prime}(G^\prime-{\partial
\bar\zeta\over\partial u})+\gamma_{AB}\lambda^A\lambda^B(\bar G^\prime-
{\partial\zeta\over\partial u}))(\bar\gamma_{C^\prime D^\prime}\bar\lambda
^{C^\prime}\bar\lambda^{D^\prime}M^{\prime a}+\gamma_{CD}\lambda^C\lambda^D
\bar M^{\prime a})$. Its contraction with $M^\prime_a$ and $N^\prime_a$,
respectively, are

$$\eqalignno{
\gamma_{AB}\lambda^A\lambda^B\bigl({\partial\zeta\over\partial u}-
   \bar G^\prime\bigr)+\bar\gamma_{A^\prime B^\prime}\bar\lambda^{
   A^\prime}\bar\lambda^{B^\prime}\bigl({\partial\bar\zeta\over\partial u}-
   G^\prime\bigr)&=-2\gamma_{AB}\lambda^AI^{\prime B}, &(4.12)\cr
-2\Vert z^e\Vert^2\bigl(H^\prime +G^\prime\bar G^\prime\bigr)&=1-\bigl(
   \gamma_{AB}\lambda^AI^{\prime B}\bigr)^2.&(4.13)\cr}
$$
\noindent
By (4.13) a point $p\in\$ $ is in $\$_+:=(\$-\{p_\pm\})\cap\Sigma_+$ iff
$\vert\gamma_{AB}\lambda^AI^{\prime B}\vert<1$ at $p$, $p\in\$_0:=(\$-\{
p_\pm\})\cap\Sigma_0$ iff $\gamma_{AB}\lambda^AI^{\prime B}=\pm1$ and
$p\in\$_-:=(\$-\{p_\pm\})\cap\Sigma_-$ iff $\vert\gamma_{AB}\lambda^AI^{
\prime B}\vert>1$. Taking into account (4.12) and (4.13) $\dot\beta^a_u$
and $z^a$ will have the form

$$\eqalignno{
\dot\beta^a_u&={{\rm i}\over2\Vert z^e\Vert}b\bigl(\bar\gamma_{C^\prime
     D^\prime}\bar\lambda^{C^\prime}\bar\lambda^{D^\prime}M^{\prime a}
     -\gamma_{CD}\lambda^C\lambda^D\bar M^{\prime a}\bigr)&(4.14)\cr
z^a&=-\Vert z^e\Vert^2\Bigl(N^{\prime a}+\bigl(H^\prime+G^\prime\bar
     G^\prime\bigr)L^a\Bigr) +{1\over2}\gamma_{RS}\lambda^RI^{\prime S}
     \bigl(\bar\gamma_{C^\prime D^\prime}\bar\lambda^{C^\prime}\bar
     \lambda^{D^\prime}M^{\prime a}+\gamma_{CD}\lambda^C\lambda^D\bar
     M^{\prime a}\bigr).&(4.15)\cr}
$$
\noindent
Suppose first that $p\in\$_\pm$. Then by means of the norm $\Vert Z^{
\prime e}\Vert$ (4.13) can be rewritten as $1-(\gamma_{AB}\lambda^A
I^{\prime B})^2=\pm{\Vert z^e\Vert^2\over\Vert Z^{\prime f}\Vert^2}$. Then
the unit normal of $\$_\pm$ in $\Sigma_\pm$ is

$$
v^{\prime a}={1\over\Vert Z^{\prime e}\Vert}\Bigl(\gamma_{CD}\lambda^C
I^{\prime D}Z^{\prime a}-{1\over2}\bigl(\bar\gamma_{R^\prime S^\prime}
\bar\lambda^{R^\prime}\bar\lambda^{S^\prime}M^{\prime a}+\gamma_{RS}
\lambda^R\lambda^S\bar M^{\prime a}\bigr)\Bigr).\eqno(4.16)
$$
\noindent
A straightforward calculation shows that $t^{\prime e}\varepsilon_{eabc}
\dot\beta^b_uz^c=b\Vert z^e\Vert v^\prime_a$. To determine the orientation
of the spacelike normals $v^{\prime a}$ on $\$_+$ and $t^{\prime a}$ on
$\$_-$ recall that the induced volume form $\varepsilon_{ab}$ on $\$ $ is
defined by $X^eY^f\varepsilon_{efab}$ for any {\it future} directed unit
timelike $X^e$ and {\it outward} directed spacelike $Y^f$ for which $X^eY_e
=0$, and observe that $v^{\prime a}$ is future directed and timelike on
$\$_-$. Thus the spacelike normals on $\$_\pm$ are {\it outward} directed,
hence $(t^{\prime a},v^{\prime a})$ and $(v^{\prime a},t^{\prime a})$ are
the proper frames representing the boost gauge on $\$_+$ and $\$_-$,
respectively, defined by $\Sigma^\prime$. Then it is easy to see that

$$
{v^\prime_aL^a\over t^\prime_bL^b}=-\gamma_{AB}\lambda^AI^{\prime B},
\eqno(4.17)
$$
\noindent
i.e. $\gamma_{AB}\lambda^AI^{\prime B}$ is completely determined by the
boost gauge on $\$_\pm$. Finally, as we saw, if $p\in\$_0$ then $\gamma
_{AB}\lambda^AI^{\prime B}=\pm 1$ there, and it is easy to show that
$t^{\prime a}=N^{\prime a}$ is an {\it ingoing} null normal to $\$_0$ if
$\gamma_{AB}\lambda^AI^{\prime B}=1$, and $t^{\prime a}$ is outgoing if
$\gamma_{AB}\lambda^AI^{\prime B}=-1$. \par

\bigskip
\bigskip

\noindent
{\lbf 5. Discussion, conclusions and remarks}\par
\bigskip
\noindent
{\bf 5.1}\par
\noindent
{}From the argumentation following eq.(4.7) it is clear that our convexity
condition implies the convexity conditions of Dougan and Mason; i.e. the
outgoing null normals are not converging {\it and} the ingoing null
normals are not diverging on $\$ $. Thus if the convexity
condition of Dougan and Mason is replaced by our (stronger) condition in
the Theorem of the introduction then the statements remain true, but in
addition the line element of $D(\Sigma)$ can be determined from the data
given on $\$ $. We would like to stress, however, that the stronger
convexity condition was used in the proof of Proposition 4.6 only to show
that the Riemann surfaces $S_u$ and their boundary $B_u$ have the simplest
possible topological structure. Nevertheless an antiholomorphic function on
a Riemann surface $S_u$ is determined by its value on the boundary $B_u$ of
$S_u$ even if $B_u$ has much more complicated structure. Thus one might be
able to determine the metric on $D(\Sigma)$ even if this convexity
condition is weakened. \par

\noindent
{\bf 5.2}\par
\noindent
Although the energy-momentum and angular momentum (and their Casimirs, the
mass and the spin) are among the most important quantities of physics, in
general relativity it is not obvious how they should be defined. For the
(quasi-local) energy-momentum the expression of Dougan and Mason seems
promising since it has a number of desirable properties [3,2,4,17]. In
particular the vanishing of the mass $m_{\$}$ is equivalent to a {\it
pp}-wave geometry, and, as a result of the present paper, the {\it pp}-wave
line element is completely encoded into the Sen geometry of $\$ $. This
implies that {\it any} physical quantity associated with a finite {\it
pp}-wave Cauchy development, e.g. the angular momentum, {\it must be
constructable} only from the two dimensional Sen geometry of $\$ $. The
simplest possible angular momentum expression obeying this requirement is
probably that of Ludvigsen and Vickers [18]. In our formalism the general
form of the (anti-self-dual part of the) Ludvigsen--Vickers type
quasi-local angular momentum is

$$
J^{{\uA}{\uB}}_{\$}:={2\over\kappa}\oint_{\$}\lambda^{\uA}_A\lambda^{\uB}_B
                     \gamma^{AB}{\rm d}\$,\eqno(5.1)
$$
\noindent
where $\lambda^{\uA}_A$ are the spinor constituents of the energy-momentum
generators. Thus actually they form a normalized basis in the space of
antiholomorphic spinor fields on $\$ $. If the spin vector is defined by
$S^{\ua}_{\$}:=\varepsilon^{\ua}{}_{{\ub}{\uc}{\ud}}P^{\ub}_{\$}J^{{\uc}
{\ud}}_{\$}$ then $g_{{\ua}{\ub}}S^{\ua}_{\$}P^{\ub}_{\$}=0$; i.e. if
$P^{\ua}_{\$}$ is timelike then $S^{\ua}_{\$}$ is zero or spacelike but
if $P^{\ua}_{\$}$ is null (i.e. $m^2_{\$}=0$) then $S^{\ua}_{\$}$ is
spacelike or proportional to $P^{\ua}_{\$}$. By $g_{{\ua}{\ub}}S^{\ua}_{\$}
S^{\ub}_{\$}=g_{{\ua}{\ub}}(J^{{\ua}{\uc}}_{\$}P_{\$ {\uc}})(J^{{\ub}{\ud}}
_{\$}P_{\$ {\ud}})-{1\over2}m^2_{\$}g_{{\ua}{\ub}}g_{{\uc}{\ud}}J^{{\ua}
{\uc}}_{\$}J^{{\ub}{\ud}}_{\$}$ the spin $S^{\ua}_{\$}$ is proportional to
the null $P^{\ua}_{\$}$ iff $J^{{\ua}{\ub}}_{\$}P_{\$ {\ub}}$ is null. \par
       For a moment suppose that $M$ is the Minkowski spacetime, $\{x^{\ua}
\}$ are Descartes coordinates, $K^{\ua}_a:=\nabla_ax^{\ua}$ are the
translations and $K^{{\ua}{\ub}}_a:=x^{\ua}K^{\ub}_a-x^{\ub}K^{\ua}_a$ are
the rotation Killing 1-forms. Let $\Sigma$ be a smooth compact spacelike
hypersurface in $M$ with smooth 2-boundary $\$:=\partial\Sigma$. If $n_a$
is the future directed unit normal to $\Sigma$ then $P^{\ua}_{\$}:=\int
_\Sigma K^{\ua}_eT^{ef}n_f{\rm d}\Sigma$ and $J^{{\ua}{\ub}}_{\$}:=\int
_\Sigma K^{{\ua}{\ub}}_eT^{ef}n_f{\rm d}\Sigma$ are well defined, depend
only on the boundary $\$ $ (and independent of $\Sigma$) and may be
interpreted as the quasi-local energy-momentum and angular momentum of the
matter fields associated with $\$ $, respectively. If $T^{ab}$ satisfies
the dominant energy condition then $P^{\ua}_{\$}$ is a future directed
nonspacelike vector and it is null iff the matter is pure radiation;
i.e. $T^{ab}=t k^ak^b$ for some nonnegative function
$t$ and constant null vector field $k^a$ (see [4]). Then for null
$P^{\ua}_{\$}$ we have $P^{\ua}_{\$}=I_\Sigma k^{\ua}$ and $J^{{\ua}{\ub}}
_{\$}P_{\$ {\ub}}=-I_\Sigma J_\Sigma P^{\ua}_{\$}$, where $k^{\ua}:=k^a
K^{\ua}_a$, $I_\Sigma:=\int_\Sigma tk^en_e{\rm d}\Sigma$ and $J_\Sigma:=
\int_\Sigma tk^en_ek_{\ub}x^{\ub}{\rm d}\Sigma$. Thus for pure radiation
$P^{\ua}_{\$}$ is null and is an eigenvector of the quasi-local angular
momentum. Since in this argumentation we have not used any specific
properties of the fields, e.g. the field equations, similar properties for
the gravitational energy-momentum and angular momentum may also be
expected. \par
      For the sake of simplicity let us consider axially symmetric {\it
pp}-wave Cauchy developments, suppose that the 2-surface $\$ $ lies in the
$v={\rm const}$ hypersurface of the canonical coordinate system, the
Killing vector $X^a$ of the axial symmetry is tangent to $\$ $ on $\$ $
and calculate the angular momentum according to (5.1). If we assume that
$D(\Sigma)$ does not admit an additional timelike Killing symmetry then
the constant null vector field $L^a$ must commute with $X^a$ and $L^aX_a=0$
must hold [19]. Then $X^a=({\partial\over\partial w})^a$, $\gamma_{AB}
\lambda^A\lambda^B=\sqrt2 \Vert z^e\Vert b^{-1}\bar\zeta=\sqrt2\Vert
z^e\Vert \exp(-{\rm i}w)$ and by $\L_X\Vert z^e\Vert^2=-L^aL^b\L_Xq_{ab}=0$
the norm $\Vert z^e\Vert$ does not depend on $w$. This implies that $J^{00}
_{\$}$ is zero, and hence $J^{{\ua}{\ub}}_{\$}P_{\$ {\ub}}=-P^{\ua}_{\$}(
J^{01}_{\$}+\bar J^{0^\prime 1^\prime}_{\$})$ and $S^{\ua}_{\$}=P^{\ua}_{\$}
{\rm i}(J^{01}_{\$}-\bar J^{0^\prime 1^\prime}_{\$})$. Thus for (axially
symmetric) {\it pp}-waves the Ludvigsen--Vickers definition (5.1) together
with the Dougan--Mason propagation law for the spinor fields $\lambda^{\uA}
_A$ yields physically reasonable results. \par
\noindent
{\bf 5.3}\par
\noindent
One of the most important principles of (classical and quantum) physics
is the locality [20]. In local quantum field theory one associates a net
of $C^*$-algebras $\{ {\cal A}(U_\alpha)\}$ of quantum observables with
every covering $\{U_\alpha\}$ of the spacetime manifold $M$, where the
subsets $U_\alpha$ are open and have compact closure; and certain axioms
are expected to hold for the net $\{ {\cal A}(U_\alpha)\}$. Although the
covering $\{ U_\alpha\}$ may otherwise be arbitrary, it seems natural to
choose the subsets $U_\alpha$ to be finite Cauchy developments ${\rm int}
D(\Sigma_\alpha)$ of finite spacelike hypesurfaces $\Sigma_\alpha$. In
fact, {\it any} spacetime admits a {\it countable} covering consisting of
such globally hyperbolic open domains; and one can construct the quasi-local
phase space of the fields and gravity. Hence one may hope to be able to
{\it construct} the quasi-local $C^*$-algebras ${\cal A}({\rm int}D(\Sigma
_\alpha))$. However in the light of the result of the present paper the
plane wave configurations both in electromagnetism and Einstein's gravity
can be specified by certain fields on the smooth 2-surfaces $\$_\alpha:=
\partial\Sigma_\alpha$; and hence the quasi-local algebra ${\cal A}({\rm
int}D(\Sigma_\alpha))$ would in fact be associated with the 2-surface
$\$_\alpha$. This result would provide a new example for the distinguished
role of 2-surfaces in fundamental physics [21,22].\par
\noindent
{\bf 5.4}\par
\noindent
Because of the nonlinearity of Einstein's equations it is difficult to
define the radiative modes of general relativity. It can be done in the
weak field approximation [23], for {\it pp}-waves and at null infinity
[24]; i.e. when the field equations become linear, and some absolute
structure (flat background, the space of anti-holomorphic spinor fields
and the universal structure, respectively) is available. It is, however, a
remarkable property of the Dougan--Mason energy-momentum $P^{\ua}_{\$}$
that it tends to the Bondi-Sachs energy-momentum ${}_{BS}P^{\ua}_{\$
_\infty}$ if $\$ $ tends to the spherical cut $\$_\infty$ of future null
infinity provided the {\it anti-holomorphic} spinor fields are used to
define $P^{\ua}_{\$}$. If however the {\it holomorphic} spinor fields are
used then in general $P^{\ua}_{\$}$ tends to infinity, whilst in
stationary spacetimes (i.e. in absence of radiation) it tends to the
Bondi-Sachs energy-momentum [17]. Therefore both at null infinity and in
the {\it pp}-wave case it is the {\it deviation} of the holomorphic and
anti-holomorphic structures of 2-spheres that characterizes the presence
of radiation. The deviation of these structures can however be defined for
generic 2-spheres in generic spacetimes too, yielding the possibility of
finding the unconstrained (i.e. radiative) modes of gravity at the
quasi-local level. \par
        A systematic and more detailed discussion of the quasi-local
energy-momentum and angular momentum ({\bf 5.2}), the quasi-local phase
space and a `quasi-local quantization' both of electromagnetism and the
{\it pp}-wave configurations of general relativity ({\bf 5.3}), and the
quasi-local radiative modes of general relativity ({\bf 5.4}) will be
published in separate papers. \par
\bigskip
\bigskip

\noindent
{\lbf Acknowledgements}\par
\bigskip
\noindent
I am greatful to Bob Wald and the University of Chicago for hospitality,
where a part of the present paper was completed. Thanks are due to Robert
Beig, J\"org Frauendiener, Joshua Goldberg, Jacek Jezierski, Lionel Mason,
James Nester, Ted Newman, Istv\'an R\'acz and Bob Wald for discussions;
and to Carlo Rovelli for pointing out ref. [21]. This work was partially
supported by the Hungarian Scientific Research Fund grants OTKA 1815,
OTKA T016246 and OTKA T017176; and also by the joint MTA--NSF grant 77-94
at the Hungarian Academy of Sciences and at the University of Chicago.\par
\bigskip
\bigskip

\noindent
{\lbf References}\par
\bigskip

\item{[1]} L.B. Szabados, Class.Quantum Grav. {\bf 11} 1833 (1994)
\item{[2]} L.B. Szabados, Class.Quantum Grav. {\bf 11} 1847 (1994)
\item{[3]} A.J. Dougan, L.J. Mason, Phys.Rev.Lett. {\bf 67} 2119 (1991)
\item{[4]} L.B. Szabados, Class.Quantum Grav. {\bf 10} 1899 (1993)
\item{[5]} M.M. Lipschutz, {\it Theory and Problems of Differential
           Geometry}, McGraw-Hill Book Co., New York 1969
\item{[6]} M. Spivak, {\it A Comprehensive Introduction to Differential
           Geometry} Vol 3, Publish or Perish Inc, Wilmington, Delaware 1975
\item{[7]} S. Kobayashi, K. Nomizu, {\it Foundations of differential
           geometry}, vols 1 and 2, Interscience: New York, 1964, 1968
\item{[8]} R. Geroch, S.-M. Perng, J.Math.Phys. {\bf 35} 4157 (1994)
\item{[9]} D. Kramer, H. Stephani, M.A.H. MacCallum, E. Herlt, {\it Exact
           solutions of Einstein's field equations}, Cambridge Univ. Press
           1980
\item{[10]} P.C. Aichelburg, Acta.Phys.Austriaca {\bf 34} 279 (1971)
\item{[11]} R. Penrose, W. Rindler, {\it Spinors and spacetime}, vol 1
            (Cambrigde Univ. Press, 1982)
\item{[12]} R. Geroch, A. Held, R. Penrose, J.Math.Phys., {\bf 14} 874 (1973)
\item{[13]} A. Sen, J.Math.Phys. {\bf 22} 1781 (1981)
\item{[14]} O. Reula, K.P. Tod, J.Math.Phys. {\bf 25} 1004 (1984)
\item{[15]} J. Frauendiener, Class.Quantum Grav. {\bf 8} 1881 (1991)
\item{[16]} J. Milnor, {\it Morse theory}, Annals of Mathematical Studies
            No 51, Princeton University Press, Princeton 1963
\item{[17]} A.J. Dougan, Class.Quantum Grav. {\bf 9} 2461 (1992)
\item{[18]} M. Ludvigsen, J.A.G. Vickers, J.Phys. {\bf A 16} 1155 (1983)
\item{[19]} L.B. Szabados, J.Math.Phys. {\bf 28} 2688 (1987)
\item{[20]} R. Haag, {\it Local Quantum Physics, Fields, Particles,
            Algebras}, Springer Verlag, 1992
\item{[21]} L. Susskind, {\it The World as a Hologram}, hep-th/9409089
\item{[22]} T. Jacobson, {\it Thermodynamics of Spacetime: The Einstein
            Equation of State}, gr-qc/9504004
\item{[23]} R.M. Wald, {\it General Relativity}, Univ. of Chicago Press,
            Chicago, 1984
\item{[24]} A. Ashtekar, {\it Asymptotic Quantization}, Bibliopolis,
            Naples 1987
\end